# Bonding nature and optical contrast of TiTe$_2$/Sb$_2$Te$_3$ phase-change heterostructure

Xudong Wang[1#], Yue Wu[1#], Yuxing Zhou[1,†], Volker L. Deringer[2], Wei Zhang[1,3]*

[1]Center for Alloy Innovation and Design (CAID), State Key Laboratory for Mechanical Behavior of Materials, Xi'an Jiaotong University, Xi'an 710049, China
[2]Department of Chemistry, Inorganic Chemistry Laboratory, University of Oxford, Oxford OX1 3QR, UK
[3]Pazhou Lab, Pengcheng National Laboratory in Guangzhou, Guangzhou 510320, China
[#]These authors contribute equally to this work.
[†]Present address: Department of Chemistry, Inorganic Chemistry Laboratory, University of Oxford, Oxford OX1 3QR, UK
*Email: wzhang0@mail.xjtu.edu.cn

**Abstract**
Chalcogenide phase-change materials (PCMs) are regarded as the leading candidate for storage-class non-volatile memory and neuro-inspired computing. Recently, using the TiTe$_2$/Sb$_2$Te$_3$ material combination, a new framework – phase-change heterostructure (PCH), has been developed and proved to effectively suppress the noise and drift in electrical resistance upon memory programming, largely reducing the inter-device variability. However, the atomic-scale structural and chemical nature of PCH remains to be fully understood. In this work, we carry out thorough *ab initio* simulations to assess the bonding characteristics of the PCH. We show that the TiTe$_2$ crystalline nanolayers do not chemically interact with the surrounding Sb$_2$Te$_3$, and are stabilized by strong covalent and electrostatic Ti–Te interactions, which create a prohibitively high barrier for atomic migration along the pulsing direction. We also find significant contrast in computed dielectric functions in the PCH, suggesting possible optical applications of this class of devices. With the more confined space and therefore constrained phase transition compared to traditional PCM devices, the recently introduced class of PCH-based devices may lead to improvements in phase-change photonic and optoelectronic applications with much lower stochasticity during programming.

**Introduction**

Phase-change memory is one of the leading candidates for future non-volatile memory applications.[1-11] Recently, a commercial persistent memory product based on 3D Xpoint technology has been successfully mass-produced. Moreover, advanced applications in beyond von Neumann architectures, such as neuro-inspired computing, have been extensively demonstrated.[12-17] Phase-change materials (PCMs), in particular, the Ge-Sb-Te (GST) alloys,[18] are the core component for these applications, exploiting the significant contrast in electrical resistivity or optical reflectivity between their crystalline and amorphous phases.[1] The fast reversible phase transition in amorphous-to-crystalline (SET) and crystalline-to-amorphous (RESET) operations is triggered by the Joule heating powered through electrical or optical pulses, corresponding to electronics and photonics applications, respectively. Notably, thanks to the recent development of advanced fabrication techniques for silicon photonics, PCMs have been incorporated with silicon waveguides, combining the unique optical properties of PCMs and the accurate control of light in a silicon photonics platform. These developments have led to promising applications in integrated non-volatile photonic memory, photonic neuro-inspired computing and optoelectronic applications.[19-27]

To meet the ever-growing storage and processing demands, further improvements and developments are desired to push the envelope of the device performance. One of the main issues that hinders the performance of phase-change memory is the high energy consumption required for device programming, especially the RESET process.[6] To overcome such issue, the concept of superlattice [28-30] or interfacial PCM (iPCM) [31-34] was introduced and developed. Instead of utilizing phase transitions between the amorphous and crystalline phases, iPCM was designed to induce crystal-to-crystal transitions by changing the stacking sequence in the GeTe/$Sb_2Te_3$ superlattice [31-34] via layer-switching through the swapped bi-layer defects.[35-40] However, evidence for thermal-based crystalline-amorphous transition was also reported in GeTe/$Sb_2Te_3$ superlattice-based devices. [41, 42] In addition, due to the chemically reactive nature of GeTe, interfacial intermixing is typically observed,[43-46] affecting the stability of GeTe/$Sb_2Te_3$ superlattices. Another major issue that limits the device performance stems from the long-distance mass transport, which can occur in the amorphization process during extensive cycling.[47, 48] Subjected to the electrostatic force induced by the electric field between two electrodes,[49] cation-like and anion-like atoms would move separately toward the two opposite electrodes. The electrostatically driven atomic migration would consequently lead to phase separation or void formation.[50] This issue would bring about either device failure in memory applications or device variability in neuro-inspired computing devices.

As reported in Ref. [51], we have designed and developed a new materials framework, which we call a phase-change heterostructure (PCH). In contrast to GeTe/$Sb_2Te_3$ superlattices in which both materials layers are PCMs with similar melting temperatures, PCH is a multilayer structure composed of alternately grown nanolayers of a switchable PCM and a stable confinement material (CM) that is of much higher chemical stability and higher melting point; thereby, the CM layers remain robust while switching the PCM layers. The first material combination was chosen as PCM=$Sb_2Te_3$ and CM=$TiTe_2$. The PCH-based devices showed a > 85% reduction in RESET energy as compared to the conventional GST-based devices configured in the same device setup.[51] The improved energy consumption was attributed to the fact that 40% of the PCH (the fraction of $TiTe_2$) was not switchable and their higher thermal resistance prevent heat loss during the RESET process. Moreover, CM layers can effectively prohibit the atomic migration of PCM during extensive operation cycles, largely reducing the cycle-to-cycle and device-to-device variability. The resistance drift issue has also been substantially improved due to the highly confined geometry for the PCM nanolayers.

The TiTe$_2$/Sb$_2$Te$_3$ heterostructure was firstly observed in a Ti$_{0.4}$Sb$_2$Te$_3$ alloy [52-55] after thermal annealing.[55] In Ref. [56], TiTe$_2$ and Sb$_2$Te$_3$ were repeatedly grown to utilize TiTe$_2$ as thermal barriers for memory switching, and MD simulations (heating from 600 to 1500 K) were reported which qualitatively confirmed the thermal stability of the TiTe$_2$ layer. The faster deposition process and the different annealing strategy employed in Ref. [56] induced some major differences in thin film structure, as compared with that reported in Ref. [51]. Although the excellent device performance has been demonstrated in Ref. [51], the atomic details of SET and RESET for PCH at room temperature remain to be further analyzed. Moreover, whether PCH can be employed for non-volatile photonic applications is yet unknown. In the present work, we carry out comprehensive ab initio molecular dynamics (AIMD) simulations for the prototype Sb$_2$Te$_3$-TiTe$_2$ PCH, in which Sb$_2$Te$_3$ and TiTe$_2$ are the PCM and CM, respectively. With realistic melt-quenched structures in hand, we systematically investigate their bonding nature and explicitly calculate the atomic migration path and the corresponding migration energy barriers. Finally, we investigate the optical contrast for PCH to unveil its potential for photonic applications.

**Computational details**
We performed DFT-based ab initio molecular dynamics (AIMD) simulations using the second-generation Car-Parrinello molecular dynamics scheme [57] implemented in the CP2K package,[58, 59] which is very efficient in enabling ab initio calculations in large scale systems ("large" here meaning several hundreds of atoms). We used the Perdew-Burke-Ernzerhof (PBE) functional [60] in combination with Goedecker pseudopotentials.[61] The time step was set to 2 fs. To evaluate the energy barriers for the atoms in the PCM layers passing through CM layers, climbing image nudged elastic band (CI-NEB) calculations were carried out to search for the minimum energy path as implemented in the CP2K package.[62, 63] Each possible migration path contains 13 images. Due to the complexity of the PCH model, in particular, the amorphous slab, the force convergence for the CI-NEB calculation was set to 0.15 eV/Å to achieve a practical compromise between convergence and accuracy. We tested a couple of threshold values in a related but simpler model, i.e. the migration of Sb atoms from atomic blocks to vacant structural gaps in hexagonal GeSb$_2$Te$_4$,[64] and showed that this threshold value was sufficiently accurate in calculating the migration barrier. We also note that a similar convergence criterion has been adopted in assessing the Ge diffusion path in defective crystalline GeTe in Ref. [65]. The Vienna Ab-initio Simulation Package (VASP)[66] was used to compute the electronic and optical properties of the relaxed crystalline and amorphous structures. We employed the projector augmented wave (PAW) method[67] with the PBE functional and set the energy cutoff to 500 eV. Chemical bonding analyses based on crystal orbital Hamilton populations (COHP) and atomic charges derived from a local projection of the electronic wavefunction were performed with the help of the LOBSTER code.[68-71] The frequency-dependent dielectric matrix was calculated within the independent-particle approximation, which was proven to be adequate to account for the optical contrast between crystalline and amorphous PCMs.[72-74] The bulk phase calculations of crystalline TiTe$_2$, crystalline and amorphous Sb$_2$Te$_3$ were made using their respective theoretical density.

**Results and discussion**

The schematic diagram of the PCH device is illustrated in Figure 1a. To study the amorphous structure, we built a supercell PCH model with 300 atoms in total, in which two TiTe$_2$ layers with 60 Ti and 120 Te atoms periodically separate the Sb$_2$Te$_3$ blocks with 48 Sb and 72 Te atoms. The amorphous structure was generated via a melt-quench protocol, similar to previous studies of PCMs.[75] The Sb$_2$Te$_3$ block in the PCH models was randomized at ~900–1300 K over 50 ps. This temperature window is above the melting point of bulk Sb$_2$Te$_3$ (~893 K),[76] but below that of bulk TiTe$_2$ (~1480 K).[77] During such high temperature amorphization process, the Sb$_2$Te$_3$ block was quickly randomized with intensive atomic movement; in strong contrast, the two TiTe$_2$ (CM) layers remained in their crystalline-like form, with the atoms vibrating around their equilibrium sites. This obvious contrast, qualitatively consistent with the simulations of Ref. [56], emphasizes the stability of the CM layers. The model was then quenched down to 300 K at the quenching rate of ~16 K/ps. After 30 ps equilibration at 300 K, the model was finally quenched down to 0 K and relaxed for further bonding analyses.

In terms of naming convention, we regard the PCH structure in the SET state as crystalline (c-) PCH, and in the RESET state as amorphous (a-) PCH, although the latter phase contains robust crystalline CM slabs. The final relaxed a-PCH structure is shown in Figure 1b. Apparently, the Sb$_2$Te$_3$ block is in amorphous phase, while the TiTe$_2$ region retains its crystalline order. To clearly differentiate the Te atoms in the a-Sb$_2$Te$_3$ and c-TiTe$_2$ regions, we label the Te atoms as "Te1" and "Te2" for Sb$_2$Te$_3$ and TiTe$_2$, respectively. In order to further quantitatively verify the structural distinction between PCM and CM layers, we carried out a $Q_4^{dot}$ analysis [78-80]. The bond order correlation parameter $Q_4^{dot}$ is a good indicator to distinguish crystalline- and amorphous-like local structures by using a rotationally invariant combination of spherical harmonic functions. Normally, large $Q_4^{dot}$ values correspond to a crystalline-like structure, while small ones indicate an amorphous-like structure.

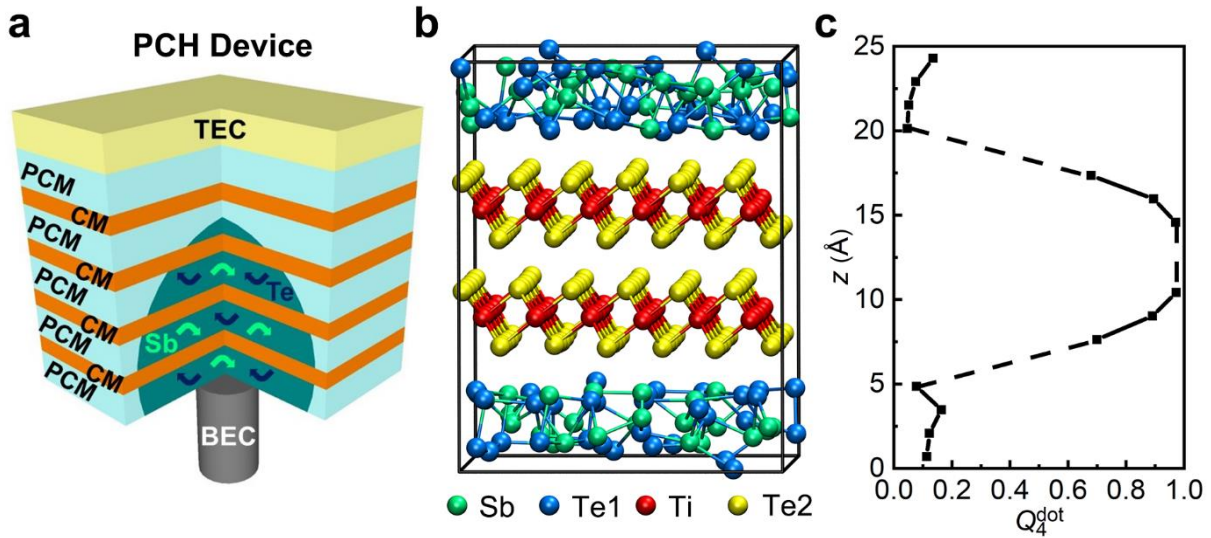

**Figure 1. The PCH framework.** (a) Schematic diagram of a typical PCH-based device; see also Ref. [51]. "TEC" and "BEC" denote top and bottom electrode contacts, respectively. (b) Atomic structure of a-PCH with two c-TiTe$_2$ layers as a CM nanolayer. Te atoms in the a-Sb$_2$Te$_3$ and c-TiTe$_2$ are labeled as Te1 and Te2, respectively. (c) The $Q_4^{dot}$ profile along the z axis. Large $Q_4^{dot}$ values represent a crystalline-like structure, while small values indicate an amorphous-like structure. Lines connecting data points are guides to the eye, and dashed lines emphasize the gap region where no atoms are found.

We divided the model into slabs in the z-direction to evaluate the averaged $Q_4^{dot}$ in each slab as shown in Figure 1c. Obviously, the plot can be separated into two major regions by clear gaps in between: one is the large $Q_4^{dot}$ region in the middle of the simulation cell coinciding with the c-TiTe$_2$ slab, and the other one is the small $Q_4^{dot}$ region on the upper and lower parts of the model (a-Sb$_2$Te$_3$ slab). The reduction of $Q_4^{dot}$ values from c-TiTe$_2$ to a-Sb$_2$Te$_3$ resembles the behavior of the interface between the amorphous and crystalline structures during the recrystallization process in the homogenous amorphous Ge$_2$Sb$_2$Te$_5$.[80] Moreover, the $Q_4^{dot}$ value for a-Sb$_2$Te$_3$ (~0.1) is also consistent with that of amorphous Ge$_2$Sb$_2$Te$_5$,[80] indicating a fully amorphized local structure for the Sb$_2$Te$_3$ region of the simulated PCH.

Next, we assessed the bonding characteristics of this PCH model using COHP analysis, which separates the covalent interactions into bonding (stabilizing; positive −COHP) and antibonding (destabilizing; negative −COHP) contributions. In our previous work,[51] we focused on the bonding features for the bulk binary phases, *viz.* crystalline Sb$_2$Te$_3$ and TiTe$_2$. In these two crystals, bonding interactions in the occupied (valence) bands were shown for TiTe$_2$, whereas non-negligible antibonding states appear just below the Fermi level for Sb$_2$Te$_3$, indicating that TiTe$_2$ is more chemically robust than Sb$_2$Te$_3$ in their respective crystalline phases. Here we studied the bonding nature in the a-PCH model, as shown in Figure 2a. A review of such methodology as applied in the PCM field is found in Ref. [81]. According to the total −COHP plot, the PCH model is overall chemically stable, with strong covalent bonding character of the occupied bands, and small yet non-negligible antibonding contributions near the Fermi level.

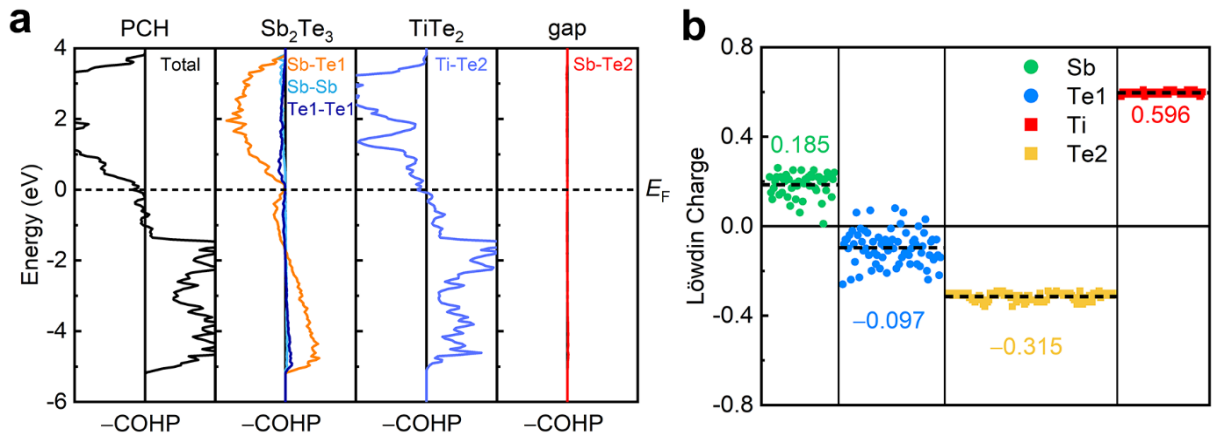

**Figure 2. Chemical bonding and charge analyses.** (a) The −COHP curve for the a-PCH model presented in Fig. 1b structure, and the projection onto the atomic contacts inside the Sb$_2$Te$_3$ and TiTe$_2$ slab, and across the gap between the two slabs. Positive or negative −COHP values represent bonding (stabilizing) or antibonding (destabilizing) interaction per simulation cell. The scale in the x-axis for the four subplots are identical. The cutoff for atomic contacts is chosen up to 3.4 Å, and five Sb-Te2 contacts are found within this range; nevertheless, no sizable covalent interaction is found across the gap region. (b) The calculated Löwdin charge for each atom in the a-PCH model. The dashed lines represent the average values for each element.

We then separated the a-Sb$_2$Te$_3$ and c-TiTe$_2$ contributions by separating the individual orbital-pair contributions to −COHP according to pairs of different neighbors. Our analysis shows that the presence of a-Sb$_2$Te$_3$ does not strongly affect the bonding nature of the TiTe$_2$ region compared to the ideal crystalline binary phase. As regards a-Sb$_2$Te$_3$, even though confined by CM, the −COHP profile is consistent with that of its bulk amorphous form,[82] implying quite similar bonding nature between PCH and the bulk counterpart. The presence of antibonding interactions in the amorphous phase is qualitatively consistent with earlier work on the amorphous PCMs.[82-88] Löwdin charges were calculated to evaluate the charge transfer and the electrostatic interactions (Figure 2b). This analysis shows that the charge transfer in c-TiTe$_2$ is much larger than that in a-Sb$_2$Te$_3$, leading to stronger electrostatic interactions. This observation is consistent with the Pauling scale of electronegativity: the values for Ti, Sb and Te atoms are 1.54, 2.05 and 2.1, respectively. The difference between Ti and Te is much larger than that between Sb and Te, leading to larger charge transfer and thus stronger electrostatic interactions. Moreover, the Löwdin charges for the a-Sb$_2$Te$_3$ and c-TiTe$_2$ regions in the simulated PCH are consistent with those of the compounds in their bulk form (the average values for Sb and Te atoms in bulk a-Sb$_2$Te$_3$ are 0.174 and -0.116, [82] and for Ti and Te atoms in bulk c-TiTe$_2$ are 0.61 and -0.31, respectively), supporting the interpretation in terms of practically negligible interactions between the a-Sb$_2$Te$_3$ region and the c-TiTe$_2$ slab. From both the covalent and electrostatic interactions perspective, the c-TiTe$_2$-like region in the PCH is more robust than that corresponding to a-Sb$_2$Te$_3$, leading to robust CM layers. The strong chemical interaction between Ti and Te is consistent with a recent experimental work, in which a Ti-Te interfacial layer between an amorphous GST thin film and a Ti electrode was formed at room temperature.[89] This shows again the importance of chemical stability of the CM layers in PCH in that the CM should be saturated and inert. In constructing PCH devices, chemically reactive elements (e.g. Ti) and alloys (e.g. GeTe) should be avoided to keep the CM layers intact.

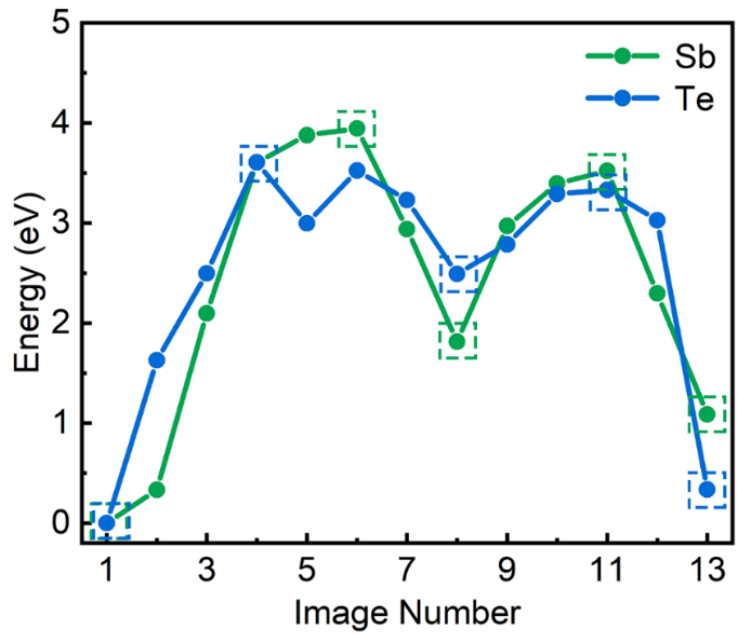

**Figure 3. Energy barriers for atomic migration.** The NEB energy profiles for one Sb and one Te atom passing through the two TiTe$_2$ layers. The barriers are calculated as 3.94 eV and 3.61 eV, respectively. The atomic images corresponding to the boxed data points are illustrated in Figure 4.

We carried out CI-NEB calculations to directly assess the potential atomic migration path for Sb/Te atom from the PCM region passing through the CM layers, mimicking possible electromigration during extensive cycling. Energy profiles for both Sb and Te atoms passing through the two TiTe$_2$ layers are shown in Figure 3. The energy barriers for Sb and Te are between 3.5 to 4.0 eV. To place these results in context, here we list a few typical energy barrier values for reference: the cation atom hopping to a nearby vacancy site in GeTe and GST compounds requires an energy value between 0.8 to 1.8 eV, depending partly on the local atomic environment;[64, 65, 90-92] the estimated amorphization energy for GST and GeTe compounds is ~2.5 and >3.0 eV respectively. [93-95] The barrier for a Sb or Te atom migrating through the CM layers is much larger (Figure 3), supporting the statement that the CM layer can indeed prohibit the atoms from long-distance migration.

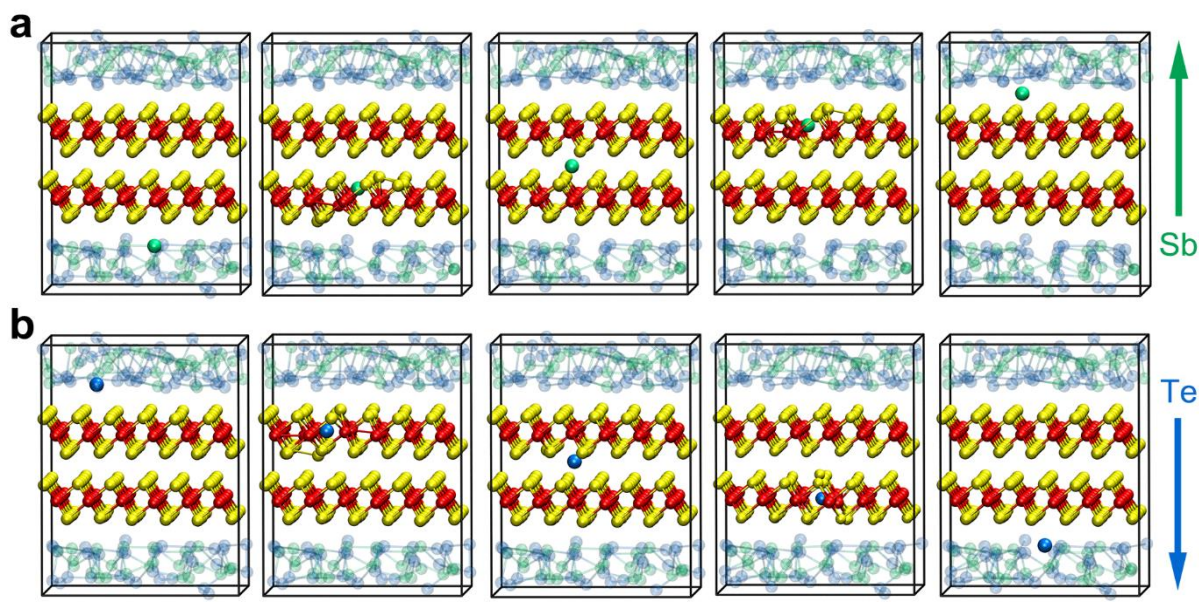

**Figure 4. Detailed migration paths.** (a) and (b) show 5 images along the NEB-computed migration path for the chosen Sb (green) and Te (blue) atom, respectively. The 5 images from left to right correspond to the initial state, the maximum energy state passing the first TiTe$_2$ layer, the intermediate saddle state residing between the two TiTe$_2$ layers, the maximum energy state passing the second TiTe$_2$ layer, and the final image.

Details of the migration process are illustrated in Figure 4. Here we picked 5 out of 13 images for each migrated atom including the initial image, the maximum energy image passing the first TiTe$_2$ layer, the intermediate image residing between the two TiTe$_2$ layers, the maximum energy image passing the second TiTe$_2$ layer, and the final image. The NEB paths for the Sb and Te atom are quite similar. Both atoms squeeze through the triangular face of a [TiTe$_6$] octahedral unit in each sublayer corresponding to the layered (CdI$_2$-type) structure of TiTe$_2$, rather than directly crossing the edge that connects the neighboring atom, minimizing the energy penalty for the migration. This pathway is consistent with the atomic migration process for GeTe and GST compounds. [64, 65, 90-92] However, since TiTe$_2$ is much denser than GST compounds, the local environment surrounding the migrated atom is so crowded that even the Te atoms of TiTe$_2$ are displaced far from their equilibrium sites, as shown in Figure 4. This strong structural distortion could account for the large energy increase during the migration. The model we built only contains two TiTe$_2$ layers due to the limitation of computational resources, whereas our previous experimental work showed that the CM layers can be easily grown around five TiTe$_2$ layers thick. [51] The larger number of CM layers would help to further reduce the probability of atomic migration compared to the idealized model considered here, and thereby further enhance the cycling performance of a PCH device. We note that defects could be present in TiTe$_2$ layers in devices, including point defects, stacking faults and grain boundaries, due to different growth and annealing conditions,[56] or potential cumulative effects upon extensive cycling; the majority of such defects will be outside the scope of DFT-based simulations. More research efforts are anticipated to assess the cycling limit and failure mechanisms of PCH-based devices.

In terms of photonic applications, most demonstrations to date have used conventional GST compounds, which implies that these devices would still suffer from the cycling and variability issues. Lately, researchers utilized pure elemental Sb in the form of few nanometer-thick layers, which could be programmed both electrically and optically.[96-99] In principle, better cycling endurance with smaller noise can be expected for the pure Sb device, because the nearly neutrally charged Sb atoms would not be displaced by the electrostatic force. It is interesting and worthwhile to test whether PCH can also be applied in optoelectronics and photonics.

To obtain an accurate prediction of the optical contrast between a- and c-PCH, we built a larger model, with a larger in-plane simulation-cell length that minimizes the lattice mismatch between the crystalline phases. The model contains 654 atoms in a supercell with lattice parameters of $a=b=26.38$ Å, $c=34.49$ Å, $\alpha=\beta=90°$, $\gamma=120°$, as shown in Figure 5a. The a-PCH structure was generated by melting the initial crystalline configuration in AIMD simulations, following by a rapid quenching down to zero K. The two models have the same volume and therefore the same mass density, relevant to the experimental condition in highly confined memory cells.

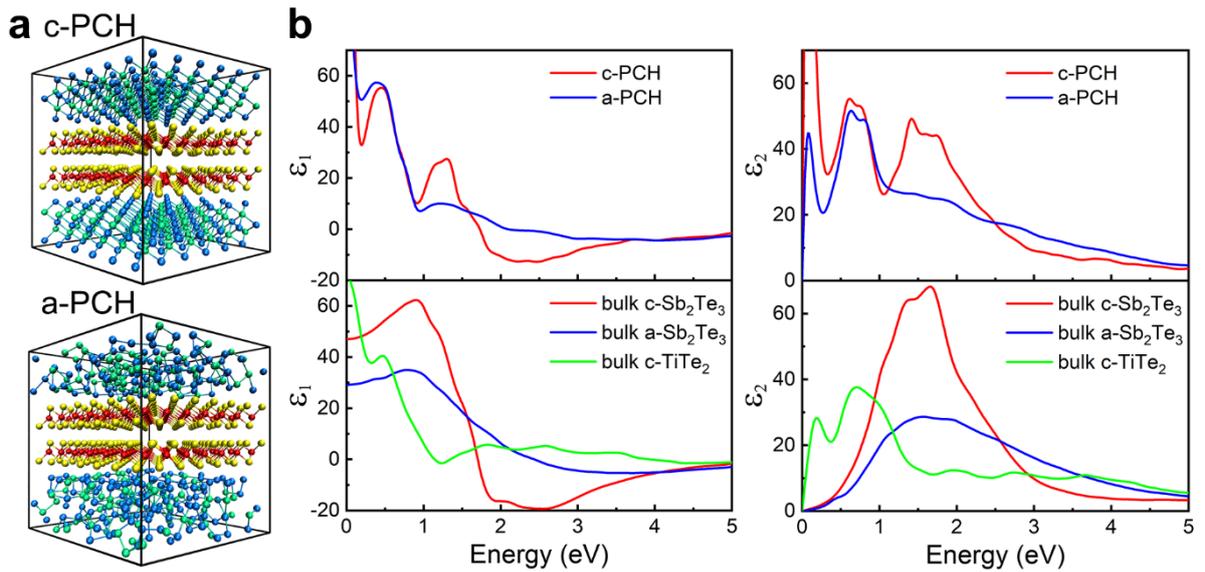

**Figure 5. Optical contrast.** (a) The atomic models of c-PCH and a-PCH used for the optical calculations. (b) The real ($\varepsilon_1$) and imaginary ($\varepsilon_2$) part of the dielectric function for c- and a-PCH (top) and the bulk c-Sb$_2$Te$_3$, a-Sb$_2$Te$_3$ and c-TiTe$_2$ (bottom).

For typical PCMs such as GeTe and GST compounds, computed dielectric functions of both amorphous and crystalline phases match well with the experimental data. [72] Figure 5b shows the real ($\varepsilon_1$) and imaginary ($\varepsilon_2$) part of the computed dielectric function of a- and c-PCH. We can see an obvious contrast between the two phases in the visible light region ranging from 1.64 to 3.19 eV. At smaller energies, the contrast is much smaller. To gain a better understanding of this behavior, we calculated $\varepsilon_1$ and $\varepsilon_2$ of the respective bulk form shown in Figure 5b. It turns out that the peak positions for $Sb_2Te_3$ and $TiTe_2$ do not overlap and they match quite well with the peak positions in PCH. Therefore, the presence of the $TiTe_2$ layers has limited effect on the $Sb_2Te_3$ region in the PCH, consistent with the fact that there is no pronounced chemical interaction between the two regions. The optical properties in the visible light region mainly stem from those of $Sb_2Te_3$, resulting in the large optical contrast for PCH.

If the tightly constrained phase transitions in PCH can also be driven by laser irradiation, the deviation in chemical composition upon cycling should be much improved, leading to much less noise and better programming consistency. Since multilevel resistance states have been realized via programming the PCH devices via multiple electrical pulses of varied pulse amplitude and duration [30], we speculate that similar partial changes in optical reflectivity or absorption could also be achieved through laser pulses. In addition to the materials screening criteria we proposed in Ref. [51], one should also avoid using a CM with absorption peaks which overlap with that of PCM for PCH-based optoelectronics and photonics. This is because the CM remains stable in both a- and c-PCH, and if there was a substantial peak overlap it would reduce the optical contrast. We note that a very recent work was published during the reviewing process of this paper, which reported the change of optical reflectivity in crystalline $TiTe_2/Sb_2Te_3$ multilayer thin films as a function of thickness variation of $Sb_2Te_3$ and $TiTe_2$ nanolayers.[100] We anticipate more experimental and theoretical efforts to address the optical property of PCH thin films and devices.

**Conclusions and outlook**
In summary, we performed systematic AIMD and DFT calculations to investigate the bonding nature and optical contrast of PCH. A clear bonding contrast exists between the different regions of the heterostructure: the chemical bonds in the CM layers are predicted to be stronger than those in the PCM layers, consistent with the observation that the crystalline CM persists during simulations at high temperature whereas the PCM quickly melts. To quantitatively evaluate the possibility of PCM atoms penetrating the CM layers, we carried out CI-NEB calculations. The computed migration barriers are quite high, amounting to >3.5 eV, even larger than the amorphization energy of GeTe or GST compounds. These large barriers effectively prevent the atoms in PCM from passing through the CM layers. Furthermore, we investigated the optical contrast of PCH and found that despite the presence of CM layers, a large optical contrast as exists between c- and a-$Sb_2Te_3$ is similarly predicted in PCH. Based on our *ab initio* simulations in Ref. [51] and in the present work, we conclude that strong chemical bonding in the CM layers is responsible for the elongated cycling endurance and improved device variability for PCH, and we suggest a possible use of $TiTe_2/Sb_2Te_3$ in optoelectronic and photonic applications.

The present study is expected to serve as a stimulus for further investigations into the atomic-scale structure and properties of PCH. Beyond the system sizes that are accessible to DFT-based MD simulations, we anticipate that machine-learned interatomic potentials [101, 102] will enable future, larger-scale modeling of structure and dynamics in PCH phases. Indeed, such potentials have been developed for GeTe [103] and $Ge_2Sb_2Te_5$,[104] and applied to bulk and nanowire structures containing several thousands of atoms [105, 106]. The addition of $TiTe_2$ (or other CM) layers will require the development of new machine-learned potentials with appropriate reference databases that cover more complex structural environments; however, the absence of substantial chemical interactions between the rigid CM layers and the PCM region, as observed in the present DFT-based simulations, indicates that the required increase in structural complexity might be relatively modest (compared to existing potentials for the pure PCMs). Moreover, whilst the present study, and that in Ref. [51], has focused on the use of $TiTe_2$ as a chemically inert CM, we suggest that different materials combinations could be screened out by high-throughput [107-109] or machine-learning [110, 111] methods, thereby enabling further developments of PCH-based devices for various demanding phase-change applications.

**Acknowledgements**

W. Z. acknowledges the National Natural Science Foundation of China (Grant No. 61774123), 111 project 2.0 (BP2018008) and the International Joint Laboratory for Micro/Nano Manufacturing and Measurement Technologies of XJTU. Y.Z. acknowledges the financial support by Chinese Scholarship Council and Oxford University. The authors acknowledge the computational resource provided by the HPC platform at XJTU.

**Conflict of Interest**

The authors declare no conflict of interest.